\documentclass[
 reprint,
 superscriptaddress,
 amsmath,amssymb,
 aps,
 floatfix,
]{revtex4-2}
\usepackage{graphicx}
\usepackage{placeins}
\usepackage{dcolumn}
\usepackage{bm}
\usepackage{hyperref}
\usepackage{xcolor}
\usepackage{soul}

\usepackage{array}
\usepackage{ragged2e}

\usepackage{tabularx}
\newcolumntype{L}[1]{>{\raggedright\arraybackslash}p{#1}}

\sethlcolor{yellow}

\begin{document}

\title{{FiuMD}: Heterogeneous Batching for Coarse-Grained Molecular Dynamics with Neural Network Potentials}

\author{Hugo D. Hern\'andez R.}
\affiliation{Instituto de Física del Sur (IFISUR), CONICET, 12 de Octubre 1865, B8000CPB Bahía Blanca, Argentina}
\affiliation{Departamento de Física, Universidad Nacional del Sur (UNS), Av. L. N. Alem 1253, B8000CPB Bahía Blanca, Argentina}
\author{Paola B. Torres}
\email[Corresponding author: ]{ptorres@frsr.utn.edu.ar}
\affiliation{Grupo Vinculado Bionanotecnología y Sistemas Complejos, INFAP (CONICET--UNSL--UTN), Av. Gral. J. J. de Urquiza 314, 5600 San Rafael, Mendoza, Argentina}
\author{Lara Negri}
\affiliation{Instituto de Física del Sur (IFISUR), CONICET, 12 de Octubre 1865, B8000CPB Bahía Blanca, Argentina}
\affiliation{Departamento de Física, Universidad Nacional del Sur (UNS), Av. L. N. Alem 1253, B8000CPB Bahía Blanca, Argentina}
\author{Rocío L. Cecchini}
\affiliation{Instituto de Ciencias e Ingeniería de la Computación (ICIC), Universidad Nacional del Sur (UNS), CONICET, San Andrés 800, 8000 Bahía Blanca, Argentina}
\author{Nicolás A. García}
\affiliation{Instituto de Física del Sur (IFISUR), CONICET, 12 de Octubre 1865, B8000CPB Bahía Blanca, Argentina}
\affiliation{Departamento de Física, Universidad Nacional del Sur (UNS), Av. L. N. Alem 1253, B8000CPB Bahía Blanca, Argentina}

\begin{abstract} %
FiuMD is a GPU molecular dynamics engine that advances many independent biomolecular simulations as a single batch, even when the systems differ in size, topology, force-field parameters, periodic box, timestep, or thermostat parameters.
All systems are concatenated into global tensors, so the classical interaction terms and a shared TorchMD-Net neural network potential are evaluated for the whole batch at once, while each system keeps its own integration and thermostat.
FiuMD started as a fork of TorchMD and reworks its system representation, force evaluation, and integration to handle such heterogeneous batches.
We test it with coarse-grained C$\alpha$ models of chignolin, Trp-cage, BBL, and $\lambda$-repressor.
In double precision, forces within a heterogeneous batch agree with isolated FiuMD and TorchMD evaluations to within the precision of repeated evaluations, with a maximum relative difference of $7\times10^{-16}$ across 60 systems. 
Over 1000 steps without a thermostat, batched trajectories deviate by at most $2.5\times10^{-11}$~\AA{} from isolated ones, even when periodic images lie within the neural network cutoff. 
With a different target temperature for each system of the ensemble, the mean and width of each system's kinetic temperature match the values expected from its own thermostat.
We show that when individual simulations are too small to fill the GPU, batching can improve throughput several-fold over TorchMD.
On a single GPU, batching four different protein systems achieves a 4.6–7.8-fold speedup per step compared with sequential TorchMD simulations.
Compared with sequential homogeneous FiuMD batches, heterogeneous batching provides up to a 4-fold speedup for smaller batches while maintaining comparable performance for larger ones. 
\end{abstract}

\maketitle

\section{Introduction}
Molecular dynamics (MD) simulations are a fundamental tool for studying the structural, dynamic, and thermodynamic properties of molecular systems, with applications in physics, chemistry, biology, and materials science.
Traditionally, these simulations have been based on classical force fields, which allow the interactions between particles to be described at a considerably lower computational cost than that of electronic structure methods, such as density functional theory (DFT) \cite{Kresse1996, Kresse1996b}.
However, the accuracy and generalizability of classical force fields can be limited, especially when attempting to describe systems or physicochemical conditions different from those for which they were parameterized.

\begin{figure*}
    \centering
    \includegraphics{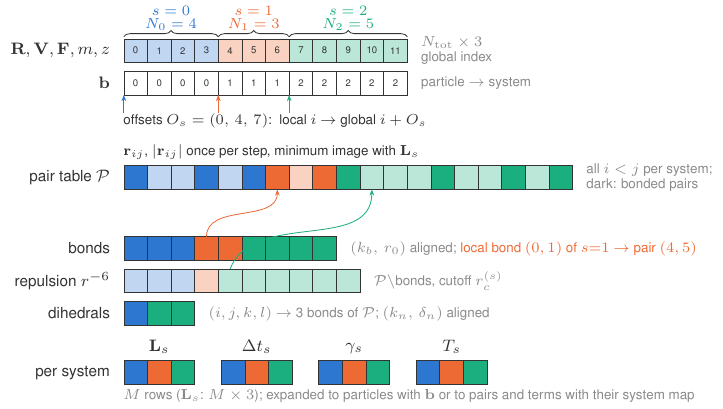}
    \caption{Global tensors of a heterogeneous ensemble.
    Toy ensemble of $M=3$ molecules with $N_s = 4$, 3, and 5 beads; colors identify each system.
    Per-particle quantities (positions $\mathbf{R}$, velocities $\mathbf{V}$, forces $\mathbf{F}$, masses $m$, and bead types $z$) are concatenated into tensors with $N_{\mathrm{tot}}$ rows.
    The batch vector $\mathbf{b}$ maps each particle to its system, and the local index $i$ of a particle of system $s$ becomes $i+O_s$.
    The pair table $\mathcal{P}$ lists every intra-system pair $i<j$; the displacement $\mathbf{r}_{ij}$ and the distance $|\mathbf{r}_{ij}|$ are computed for all of them once per step, with the minimum image in the box $\mathbf{L}_s$ of the system of the pair.
    Bonds are stored as indices into $\mathcal{P}$, dihedrals as indices of their three bonded pairs, and the repulsive term acts on the pairs of $\mathcal{P}$ that are not bonds.
    Each list has a parameter tensor of the same length.
    The lists have a different length for each system; here the second system has no dihedral.
    Box lengths, timestep, friction coefficient, and target temperature are stored with one row per system.
    Arrows show one bond and one repulsive pair pointing into $\mathcal{P}$.}
    \label{fig:layout}
\end{figure*}

In recent years, machine learning (ML) methods have played an increasingly important role in the development of molecular interaction potentials.
In particular, neural network potentials (NNPs) \cite{behler2007, botu2017, deringer2019, kocer2022, schnet} allow for the high-precision approximation of potential energy surfaces, while also offering the ability to capture complex relationships between molecular configurations and their properties \cite{batatia2022, yang2024, park2024, schnet, gao2020, zeng2023, torchmdnet}.
Among these approaches, message-passing neural networks (MPNNs) allow for the construction of representations of molecular environments through the iterative exchange of information between particles \cite{gilmer2017}.

At the same time, coarse-grained (CG) models make it possible to extend the spatial and temporal scales accessible through molecular simulations.
To achieve this, they reduce the level of detail in the molecular representation by grouping several atoms into effective units (beads).
While this simplification reduces computational cost, it also poses the challenge of developing interaction potentials capable of reproducing the relevant properties of the original system.
In this context, CG NNPs constructed from atomistic information using bottom-up strategies represent a promising alternative.
In particular, $\Delta$-learning approaches combine a reference physical potential with a correction learned via neural networks, allowing researchers to leverage the knowledge embedded in the classical model and improve its predictive capacity.
This type of strategy has proven effective for developing CG potentials that are transferable across different protein systems \cite{majewski2023}.
Other recent developments have explored neural CG models capable of generalizing to proteins and sequences different from those used during training \cite{charron2025}.

The development of these models has driven the creation of computational libraries that facilitate both their implementation and their integration into molecular simulations.
Among these are SchNetPack \cite{schnet}, TorchANI \cite{gao2020}, and DeePMD-kit \cite{zeng2023}.
Other tools that directly integrate molecular dynamics with machine learning frameworks have also emerged, such as JAX MD \cite{jaxmd} and TorchMD \cite{torchmd}.
In particular, TorchMD provides a PyTorch-based \cite{pytorch} infrastructure that enables GPU-accelerated molecular simulations and integrates neural potential models into molecular dynamics workflows.
Its flexibility allows for the combination of machine learning models with traditional components of molecular simulations, laying the groundwork for the development of new simulation methodologies.

However, efficiently running molecular simulations on GPUs presents particular challenges when working with small systems.
In these cases, the computational work associated with an individual simulation may be insufficient to fully utilize the available GPU resources, while kernel-launch and CPU--GPU communication overheads can represent a significant fraction of the total execution time.
This limitation is particularly relevant for coarse-grained models of small proteins, which constitute one of the systems of interest in this work.
Processing multiple independent simulations simultaneously is therefore a natural strategy to increase the amount of computational work performed in parallel.
Although existing libraries provide mechanisms for batching molecular configurations, their implementations differ in the types of systems that can be processed together.
For instance, TorchMD propagates several replicas of the same molecular system as one batch~\cite{torchmd}, and mlcg provides tools for batched simulations of coarse-grained models~\cite{charron2025}.
In TorchMD the replicas share one topology and one set of force-field parameters, so molecules with different numbers of particles or different topologies cannot be combined within the same batch.

\begin{table*}[t]
\caption{\label{tab:scope}Scope of a batch in FiuMD v0.1.0~\cite{fiumd_v010}. The second column states whether a setting can differ between the systems of one batch; the third gives what the tests of Sec.~\ref{sec:tests} used. Entries marked ``not varied'' are implemented but were not exercised by any test reported here.}
\begin{ruledtabular}
\begin{tabular}{L{0.5\columnwidth}L{0.4\columnwidth}L{0.9\columnwidth}}
Setting & Per system & Used in this paper \\
\hline
Number of beads & yes & 10, 20, 47, 80 \\
Bond and dihedral lists, masses, prior parameters & yes & native sets of the four proteins \\
Repulsion coefficients by bead type & yes & native tables; no deliberate conflict between type names \\
Repulsion cutoff & yes, or none & not varied \\
Periodic box (three fixed side lengths) & yes & a different box per protein, four box sets \\
Timestep & yes & 1 and 0.5~fs in one batch \\
Target temperature & yes & four targets in one batch \\
Friction coefficient & yes & not varied (1~ps$^{-1}$) \\
Set of classical force terms & shared; bond term required & bonds, proper dihedrals, $r^{-6}$ repulsion \\

NNP checkpoint and bead-ID vocabulary & shared & $\lambda$-repressor checkpoint \\
Device and precision & shared & double (forces, trajectories); single (ladder, timing) \\
\end{tabular}
\end{ruledtabular}
\end{table*}

An alternative strategy is to represent multiple systems of potentially different sizes using concatenated global tensors.
TorchSim, for example, implements a PyTorch-based simulation engine in which particle attributes from different systems are concatenated into global tensors, allowing systems with different numbers of particles to be processed within a common computational framework \cite{torchsim}.
This approach addresses an important limitation of conventional batching strategies, since the systems no longer need to share the same particle dimensions.
Nevertheless, accommodating different particle numbers is only one aspect of heterogeneous molecular systems.
Biomolecular systems also require the explicit representation of molecular connectivity and the associated interaction parameters.
In particular, protein models may involve multiple types of bonded interactions, such as bonds, angles, and dihedrals, together with non-bonded interactions whose evaluation may depend on molecular connectivity and interaction-specific parameters.
Supporting systems with different topologies therefore requires not only concatenating particle coordinates, but also maintaining independent interaction lists and force-field parameters for each molecular system.
TorchSim concatenates systems of different sizes for atomistic simulations with machine-learned potentials; FiuMD applies concatenated tensors to the bonded topologies and per-system force-field parameters of the coarse-grained protein models used with TorchMD.

The ability to simultaneously simulate heterogeneous molecular systems is relevant to several applications in computational biophysics and molecular modeling.
For example, comparative studies involving different proteins or sequence variants require the evaluation of molecular systems with potentially different numbers of particles and molecular connectivities.
Processing these systems within a common simulation framework can facilitate direct comparisons of their structural and dynamical properties under equivalent computational conditions.
Similarly, investigating the dependence of molecular behavior on temperature or other simulation parameters may involve running independent systems under different thermodynamic conditions.
A heterogeneous system framework can accommodate these differences within a single computational workflow, avoiding the need to execute each system separately.
This capability is also particularly relevant for assessing the transferability of coarse-grained NNPs.
When a model is intended to generalize across different proteins or sequences, its performance must be evaluated on molecular systems that may differ in size, composition, and topology.
Simulating these systems within a common framework can facilitate systematic comparisons and, as shown below, reduces the wall time per step when the individual batches are small.

In this paper, we present FiuMD, a fork of TorchMD~\cite{torchmd} designed for the simultaneous simulation of multiple independent and heterogeneous biomolecular systems.
Its main contribution is to combine the flexibility of variable-size system representations with the ability to preserve and evaluate the distinct molecular topologies and interaction parameters of each system within a shared computational workflow.
FiuMD achieves this through a global tensor representation that concatenates particle coordinates, velocities, interaction indices, and force-field parameters from all systems, while maintaining their individual molecular identities and dynamical independence.
On this representation, vectorized force evaluation and time integration algorithms enable the simultaneous processing of classical and neural network contributions, including systems with different thermodynamic conditions and simulation parameters.
We test the numerical equivalence of heterogeneous batches with isolated FiuMD and TorchMD runs in double precision, for forces, energies, and deterministic trajectories.
We check the kinetic-temperature distribution when the systems of one batch have different target temperatures.
Finally, we time a mixed job of four proteins against the same systems run sequentially, either as four homogeneous FiuMD batches or as four TorchMD runs.

\section{FEATURES OF FIUMD}

The core objective of FiuMD is the efficient execution of multiple independent molecular systems within a single molecular dynamics simulation workflow.
To achieve this, all systems are represented using global tensors that store particle coordinates, velocities, interaction indices, and force-field parameters.
Force evaluation and time integration are subsequently performed through vectorized tensor operations acting on the complete ensemble.

\begin{figure}
    \centering
    \includegraphics{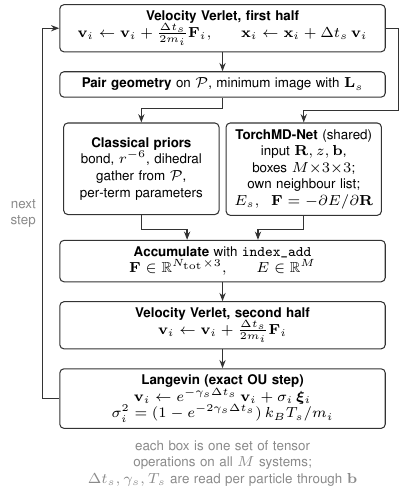}
    \caption{One step of the whole ensemble.
    The classical terms read the pair table of Fig.~\ref{fig:layout}; TorchMD-Net receives the positions, bead types, batch vector, and diagonal $M\times3\times3$ boxes, and builds its own neighbor list.
    Energies and forces are accumulated per system with indexed additions.
    The Langevin step follows the complete velocity Verlet step.
    Each box is one set of tensor operations on all $M$ systems, so the number of calls per step does not depend on $M$.}
    \label{fig:step}
\end{figure}

The implementation consists of five main components: (i) a representation of heterogeneous molecular systems, (ii) global indexing of molecular interactions, and (iii) vectorized force evaluation, (iv) neural network potential evaluation, and (v) molecular dynamics integration.
Table~\ref{tab:scope} lists which settings may differ between systems and which of them the tests below varied.

\subsection{Representation of heterogeneous systems}

Consider an ensemble composed of $M$ independent molecular systems, indexed by $s = 0,\ldots,M-1$.
System $s$ contains $N_s$ particles, such that the total number of particles in the ensemble is

\begin{equation}
N_{\mathrm{tot}} = \sum_{s=0}^{M-1} N_s.
\end{equation}

Rather than storing each system separately, FiuMD arranges all particle information in a single set of global tensors.
Particle coordinates, velocities, masses, and other atomic properties are concatenated along the particle dimension, producing arrays of size $N_{\mathrm{tot}}$.
For example, particle coordinates are represented as

\begin{equation}
\mathbf{R} \in \mathbb{R}^{N_{\mathrm{tot}} \times 3},
\end{equation}

where each row corresponds to the Cartesian coordinates of a particle belonging to one of the systems in the ensemble.

To preserve the identity of each particle, an additional system-index array is constructed.
This array maps every particle to its corresponding molecular system and enables the selective application of operations such as force accumulation, thermodynamic parameter assignment, and trajectory analysis.

The same strategy is applied to molecular interactions.
Bonded and non-bonded interactions from all systems are stored in system-wide interaction lists defined in terms of particle indices within the concatenated coordinate array.
Consequently, force calculations can be performed over the entire ensemble using a single set of tensor operations, regardless of the number of particles or interactions present in each system.

This representation allows systems with different sizes, topologies, force-field parameters, and thermodynamic conditions to coexist within the same simulation batch while maintaining complete dynamical independence.
Interactions are only defined between particles belonging to the same molecular system, ensuring that no artificial coupling is introduced by the system-wide representation.

\subsection{Global indexing of molecular interactions}

Each molecular system is initially defined using local particle indices.
For example, a bond interaction is specified by a pair of particle indices $(i,j)$, while a dihedral interaction is specified by a quadruplet $(i,j,k,l)$, where all indices refer to particles within a single system.

To construct a heterogeneous system, these local interaction definitions are transformed into a global indexing scheme.
Let
\begin{equation}
O_s = \sum_{m=0}^{s-1} N_m
\end{equation}
denote the particle offset associated with system $s$, so that $O_0 = 0$.
Any local particle index $i$ belonging to that system is mapped to a global index
\begin{equation}
i_{\mathrm{global}} = i + O_s.
\end{equation}
The same transformation is applied to every interaction in the system.
As a result, bonded and non-bonded interactions from all molecular systems can be represented using a single collection of global index tensors.

\begin{table*}[t]
  \caption{ Coarse-grained systems. One bead per residue (C$\alpha$), mass 12~u, one bead type per amino acid. 
  All four proteins use the same classical terms: harmonic bonds between consecutive beads, a two-term periodic dihedral ($n=1,2$) on every four consecutive beads, and a repulsive $\varepsilon_{ij}/r^{6}$ term on every intra-system pair that is not bonded with no cutoff.
  Box side lengths are in~\AA{}; the data boxes are those of the reference trajectories}
  \label{tab:systems}
  \begin{ruledtabular}
  \begin{tabular}{lcccccccccc}
  Protein & Beads & Bonds & Dihedrals & Repulsive & \multicolumn{4}{c}{Box side (\AA)} & $\Delta t$ (fs) & $\Delta t$ (fs)\\
          & $N_s$ & & & pairs & data & large & tight & images & static/NVE/ladder & timing\\
  \hline
  Chignolin (CLN)        & 10 &  9 &  7 &   36 & $\approx40$ & 30 & 25 & 24 & 1   & 1\\
  Trp-cage (TRP)         & 20 & 19 & 17 &  171 & $\approx38$ & 40 & 26 & 24 & 1   & 1\\
  BBL                    & 47 & 46 & 44 & 1035 & $\approx60$ & 55 & 28 & 31 & 0.5 & 1\\
  $\lambda$-repressor    & 80 & 79 & 77 & 3081 & $\approx71$ & 70 & 30 & 39 & 0.5 & 1\\
  \end{tabular}
  \end{ruledtabular}
\end{table*}

For example, all bond interactions are stored in an array
\begin{equation}
\mathbf{B}
=
\left[
(i_1,j_1),
(i_2,j_2),
\ldots,
(i_{N_b},j_{N_b})
\right],
\end{equation}
where $N_b$ denotes the total number of bonds across all systems and each index refers directly to particles in the global coordinate tensor.
Analogously, dihedral interactions are stored as quadruplets of global particle indices.

Interaction-specific parameters are concatenated using the same procedure.
Equilibrium bond lengths, force constants, dihedral coefficients, and other force-field parameters are assembled into global parameter tensors whose elements correspond one-to-one with the interaction lists.
Consequently, each interaction is self-contained within the system-wide lists and can be evaluated independently of its molecular origin.

The resulting representation transforms a collection of heterogeneous molecular systems into a unified set of particle, interaction, and parameter tensors.
This structure enables subsequent force calculations to be expressed entirely through vectorized tensor operations acting on the complete ensemble.

\subsection{Vectorized force evaluation}

The force calculation in FiuMD is formulated as a sequence of vectorized tensor operations over the complete ensemble.
Each force contribution is evaluated independently using the global particle indices and interaction parameters described above, and the resulting forces are accumulated into a single global force tensor,

\begin{equation}
\mathbf{F} \in \mathbb{R}^{N_{\mathrm{tot}} \times 3}.
\end{equation}

For a given interaction type, particle coordinates are gathered from the global coordinate tensor using the corresponding interaction indices.
Geometric quantities required by the force expression are then evaluated simultaneously for all interactions.
For example, for a set of two-body interactions between particles $i$ and $j$, the displacement vectors are obtained as

\begin{equation}
\mathbf{r}_{ij}
=
\mathbf{R}_{j}-\mathbf{R}_{i},
\end{equation}

where the indices $i$ and $j$ are taken from the global interaction tensor.
The resulting collection of displacement vectors is processed simultaneously using PyTorch tensor operations, rather than iterating over individual molecular systems.

The forces generated by each interaction are subsequently accumulated onto the corresponding particles.
Since multiple interactions may contribute to the force acting on the same particle, this accumulation is performed using indexed tensor operations.
The resulting global force tensor therefore contains the total force acting on every particle in the ensemble while preserving the independence of the individual molecular systems.

The same strategy is used for the three classical terms currently vectorized in FiuMD: bond, repulsive $r^{-6}$, and dihedral interactions.
No angle term is implemented.
The two-body terms (bonds and repulsion) are evaluated from globally indexed particle pairs, and the dihedral term is evaluated from globally indexed particle quadruplets.
Interaction parameters are stored in tensors aligned with their corresponding interaction lists, allowing the parameters of different molecular systems to be processed in the same operation even when they have different values.

Periodic boundary conditions (PBCs) are applied directly to the displacement vectors prior to force evaluation.
The classical terms use the minimum-image convention for orthorhombic boxes,
\begin{equation}
\mathbf{r}_{ij} \leftarrow \mathbf{r}_{ij} - \mathbf{L}_s \operatorname{round}\left(\frac{\mathbf{r}_{ij}}{\mathbf{L}_s}\right),
\end{equation}
applied componentwise, where $\mathbf{L}_s$ contains the three box lengths of the system $s$ to which the pair belongs.
The box lengths of all systems are stored in a tensor of shape $M\times3$, so systems within an ensemble may have different box dimensions.

For NNPs, PBCs are handled internally by the NNP implementation.
FiuMD provides the simulation box together with the global particle coordinates and system assignment vector.
The per-system box lengths are passed as diagonal box matrices in a tensor of shape $M\times3\times3$.
The NNP distance calculation then associates each particle with the corresponding molecular system through the batch vector and applies the appropriate periodic geometry during neighbor-list construction.

This separation allows the classical force implementation to remain independent of the NNP-specific neighbor-list and periodic-distance machinery, while still supporting independent simulation boxes for the molecular systems contained in an ensemble.

\begin{table*}[t]
  \caption{Simulation protocols. All runs: FiuMD v0.1.0~\cite{fiumd_v010}, PyTorch~2.13.0, TorchMD~1.1.2, TorchMD-Net~3.0.3, the $\lambda$-repressor NNP checkpoint for every protein, one NVIDIA GeForce RTX~5060~Ti. Composition is given as CLN/TRP/BBL/$\lambda$-repressor replicas. Langevin runs apply the Ornstein--Uhlenbeck velocity update after the complete velocity Verlet step.
  }
  \label{tab:protocols}
  \begin{ruledtabular}
  \setlength{\tabcolsep}{3pt}\footnotesize
  \begin{tabular}{L{2.1cm}L{1.0cm}L{3.3cm}L{1.1cm}L{3.2cm}L{2.6cm}L{2.0cm}}
  Protocol & Shown in & Systems (CLN/TRP/BBL/$\lambda$) & Precision & Dynamics & Length & Measured\\
  \hline
  Static evaluation & Fig.~\ref{fig:forces} & 60 (32/16/8/4) & float64 & one force call & 1 evaluation per box set (data, large, tight) & $\mathbf{F}$, $E$ vs.\ 5 references\\[2pt]
  Deterministic & Fig.~\ref{fig:nve} Fig~\ref{fig:nve_time} & 60 (32/16/8/4), large boxes; 31 (3/16/8/4), image boxes & float64 & velocity Verlet, no thermostat; initial velocities at 300/350/370/400~K & 1000 steps & max.\ bead separation\\[2pt]
  Temperature ladder & Fig.~\ref{fig:temperature} & FiuMD: 4 batches of 36 (16/8/8/4), one target per protein in each batch; TorchMD: 4 runs of 24 (8/8/4/4), one target per run & float32 & Langevin, $\gamma=1$~ps$^{-1}$, $T=300$, 340, 380, 420~K & $3\times10^{5}$ steps, one frame every $10^{3}$ & $\langle T\rangle$, $\sigma_T$%
  \\[2pt]
  Timing & Fig.~\ref{fig:timing} & $M=4$--512, $M/4$ of each & float32 & Langevin, $\gamma=1$~ps$^{-1}$, $T=350$~K & 20 warm-up $+$ 5$\times$100 steps & wall time, peak memory\\
  \end{tabular}
  \end{ruledtabular}

\end{table*}

Importantly, no neighbor-list construction is performed for the classical force terms.
The displacement vectors of all $N_s(N_s-1)/2$ particle pairs within each system are evaluated at every step, so the cost of the classical terms grows quadratically with system size.
The repulsive term acts on every intra-system pair that is not directly bonded and can be truncated at a cutoff specified per system.
This choice is appropriate for the intended coarse-grained simulations because the computational cost of the neural network potential is substantially larger than that of the classical force terms.
Furthermore, the neural network potential performs its own neighbor-list construction internally.

Because all classical force contributions are evaluated on the global tensors, the computational cost associated with iterating over individual systems is removed from the force-calculation path.
The resulting force evaluation is independent of the number of systems in the ensemble at the Python level, allowing the GPU to process interactions from many independent molecular systems concurrently.

\subsection{Neural network potential evaluation}

NNPs are evaluated on the complete ensemble using a single batched model evaluation.
The particle coordinates of all molecular systems are provided to the NNP as a single tensor, while an additional batch index identifies the molecular system to which each particle belongs.
This allows the NNP implementation to process multiple independent molecular systems concurrently without requiring the systems to have the same number of particles or topology.

For an ensemble containing $N_{\mathrm{tot}}$ particles, the batch vector is represented by
\begin{equation}
\mathbf{b} \in \{0,\ldots,M-1\}^{N_{\mathrm{tot}}}
\end{equation}
and it assigns each particle to one of the $M$ molecular systems.
The batch information is passed to the NNP together with the particle positions and atomic embeddings, allowing the neural network to distinguish the independent systems within the global representation.

The NNP is evaluated once for the complete ensemble, producing an energy for each molecular system.
The corresponding forces are obtained from the negative gradient of the total energy with respect to the particle coordinates,

\begin{equation}
\mathbf{F}_{\mathrm{NNP}}
=
-\frac{\partial E}{\partial \mathbf{R}}.
\end{equation}

In the implementation, the NNP parameters are kept fixed during molecular dynamics and gradients are computed only with respect to the particle positions.
The forces are therefore obtained through automatic differentiation of the NNP energy with respect to the coordinates.

Neighbor-list construction for the neural interactions is handled internally by the NNP implementation.
The batch assignments ensure that particles belonging to different molecular systems are treated as separate systems when constructing the neural interactions.
FiuMD therefore does not explicitly construct or maintain a separate neighbor list for the NNP.

The resulting NNP energies and forces are accumulated directly into the global ensemble energy and force tensors.
Because the NNP evaluation operates on the concatenated representation, the computational work associated with multiple independent molecular systems can be submitted to the accelerator as a single batched operation.
This avoids performing separate NNP evaluations for each system and allows the computational resources of the accelerator to be shared across the ensemble.

\begin{figure*}
    \centering
    \includegraphics[width=0.8\textwidth]{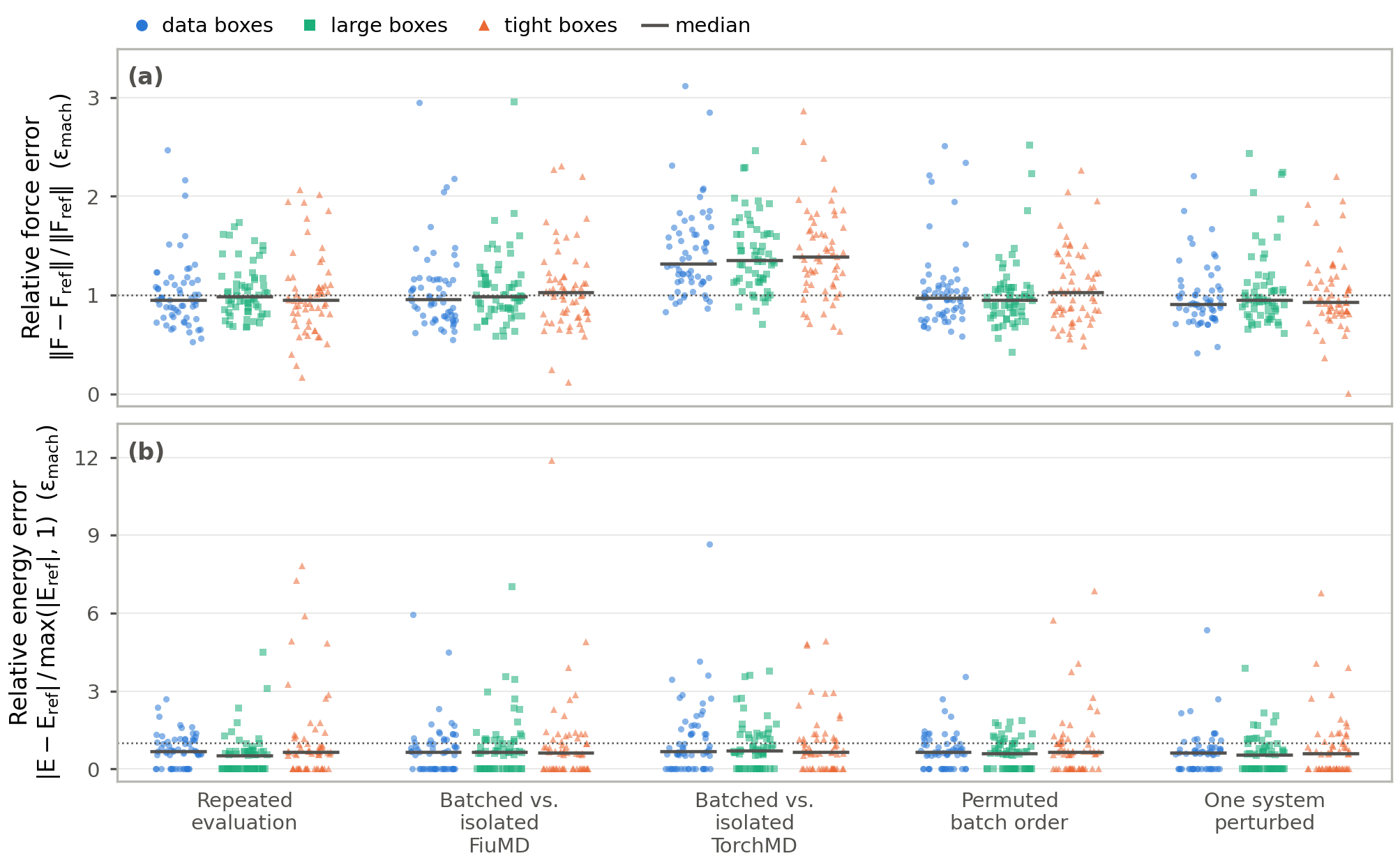}
    \caption{Force and energy errors of a heterogeneous batch, in units of the double-precision machine epsilon $\epsilon_{\mathrm{mach}} \approx 2.2\times10^{-16}$. The batch holds 60 systems (32 chignolin, 16 Trp-cage, 8 BBL and 4 $\lambda$-repressor) evaluated with the priors and the neural network potential. (a)~Relative error of the forces of each system.(b)~Relative error of its potential energy. Each point is one system and each dash the median of its box set, see Tab~\ref{tab:systems}.    
    The columns compare the batch with a second evaluation of itself, with each system evaluated alone in FiuMD and in TorchMD, with the same batch in permuted order, and with the batch after the positions of one system were perturbed (the other 59 systems are shown). The dotted line marks $\epsilon_{\mathrm{mach}}$.}
    \label{fig:forces}
\end{figure*}

All molecular systems within a given ensemble use the same NNP architecture and model parameters.
The systems may nevertheless differ in particle number, atomic configuration, and topology, provided that they are compatible with the input representation expected by the NNP.

\subsection{Molecular dynamics integration}

Time integration in FiuMD is implemented through modular PyTorch-based integrator classes operating directly on the global ensemble tensors.
The integrators update the positions and velocities of all particles simultaneously, while system-specific simulation parameters are stored according to the molecular system to which each particle belongs.
This design allows independent molecular systems to be propagated within the same ensemble while retaining distinct integration parameters.
Figure~\ref{fig:step} summarizes one step of the whole ensemble.

For deterministic molecular dynamics, FiuMD implements the velocity Verlet algorithm, for which the timestep is specified independently for each molecular system and expanded to the corresponding particles, allowing systems with different integration timesteps to be propagated within the same ensemble.
The implementation performs all operations using vectorized tensor arithmetic over all particles in the ensemble.

Temperature-controlled simulations are supported through a Langevin thermostat, implemented as a separate module that updates the velocity of each particle $i$ of system $s$ with an exact Ornstein--Uhlenbeck step,
\begin{equation}
\mathbf{v}_i \leftarrow e^{-\gamma_s \Delta t_s}\,\mathbf{v}_i + \sqrt{\left(1-e^{-2\gamma_s \Delta t_s}\right)\frac{k_B T_s}{m_i}}\;\boldsymbol{\xi}_i,
\end{equation}
where $\boldsymbol{\xi}_i$ is a vector of independent standard normal variates and $m_i$ is the particle mass.
Each FiuMD step consisted of a full velocity Verlet step followed by this thermostat update. In the TorchMD reference, the Langevin update is applied within the step, after the force evaluation and before the second velocity half-step.

FiuMD stores the friction coefficient $\gamma_s$, the target temperature $T_s$, and the timestep $\Delta t_s$ on a per-system basis and expands them to the corresponding particles in the global ensemble representation.
Consequently, different molecular systems can be simulated with different thermostat parameters while sharing the same vectorized thermostat operation.

\begin{figure}
    \centering
    \includegraphics[width=\columnwidth]{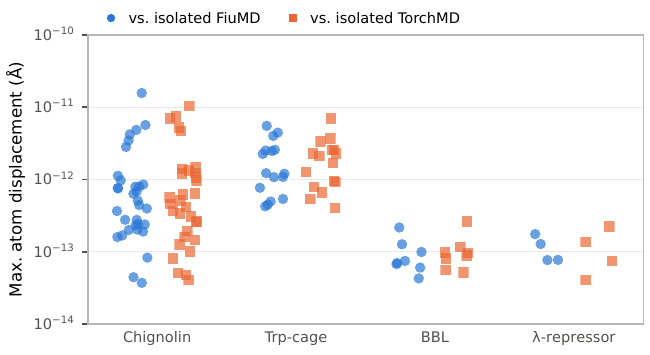}
    \caption{Divergence between batched and isolated NVE trajectories after
1000 steps. Each point is one system: the largest displacement of
any of its atoms between the batched run and the same system run alone, with
identical initial positions and velocities, in FiuMD (circles) and in
TorchMD (squares). A batch of 60 systems (32 chignolin, 16 Trp-cage, 8 BBL and
4 $\lambda$-repressor) in large boxes, integrated
in double precision with the priors and the neural network potential, see Tab.~\ref{tab:systems}. }
    \label{fig:nve}
\end{figure}

Both integrator types operate on the forces generated by the complete force-evaluation pipeline described above.
The integrator therefore does not require separate treatment of classical and neural network forces: all contributions are accumulated into the global force tensor before the integration step.
Similarly, the thermostat acts on the global velocity tensor and does not introduce coupling between molecular systems.

The integrator architecture is extensible through a common base class, allowing additional integration schemes or thermostats to be implemented while retaining the same ensemble representation.
The use of preallocated tensors for intermediate quantities also minimizes temporary memory allocation during repeated integration steps and is compatible with GPU-accelerated execution.

\section{Simulation setup}

\subsection{Coarse-grained protein systems}

Simulations were performed on coarse-grained protein models previously introduced by Majewski \textit{et al.}~\cite{majewski2023}.
In this representation, each amino acid residue is represented by a single bead, resulting in compact systems suitable for neural network potential evaluation.
Four proteins were used: chignolin (CLN, 10 beads), Trp-cage (TRP, 20 beads), BBL (47 beads), and $\lambda$-repressor (80 beads) with the parameters indicated in Table \ref{tab:systems}.

The table shows the three different box sizes for each protein. The column ``data'' depict the boxes of the reference trajectories of Majewski \textit{et al.}, which are not exactly cubic. Then, column ``Box Side'' shows three different sizes: ``large'' are loose cubic boxes; ``tight'' are small boxes used as a negative control; and ``images'' are the boxes of the second deterministic batch. The column $\Delta t$ (fs) shows the timestep used in each case for evaluated systems. 

\begin{figure}
    \centering
    \includegraphics[width=\columnwidth]{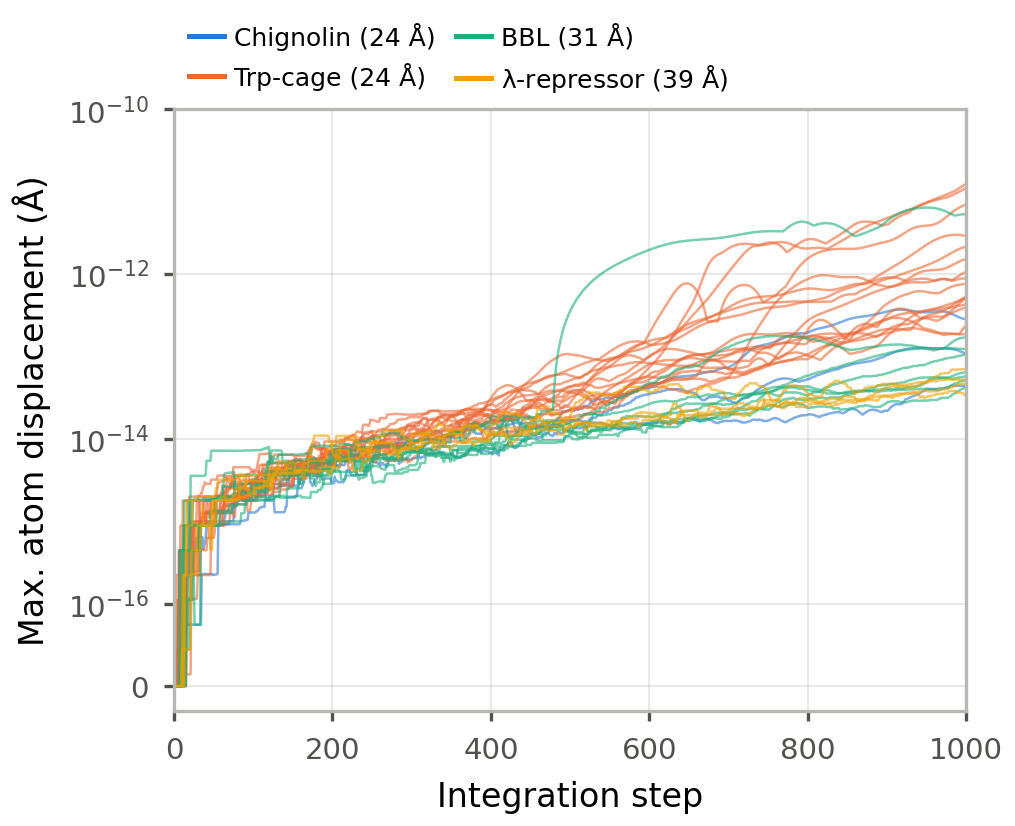}
    \caption{Divergence between batched and isolated NVE trajectories in boxes
            where each system interacts with its own periodic images. 
            Each line is one system: the largest displacement of any of its atoms between the batched run and the same system run alone in FiuMD, from identical initial conditions. 
            Points at are steps where both trajectories agree to the last bit.
            The systems were integrated for 1000 steps in double precision with the priors and the neural network potential, using $\Delta t = 1$~fs for chignolin and Trp-cage and 0.5~fs for BBL and $\lambda$-repressor.
            All systems stay below  $10^{-8}$~\AA{}, tolerance of the test.
            Must-fail controls, the same run with a wrong time step and without periodic images, diverge by $8\times10^{-3}$ and $8\times10^{-5}$~\AA{} after a single step.
            }
\label{fig:nve_time}
\end{figure}

\subsection{Classical coarse-grained interactions}

The baseline coarse-grained topology included bonded and non-bonded interactions consistent with the reference coarse-grained model.
Specifically, the topology contained bond and dihedral interactions and a purely repulsive $r^{-6}$ potential for excluded-volume interactions, applied to all pairs of beads that are not directly bonded.
In all reported runs, the repulsion was evaluated without a cutoff in both engines.

\subsection{Neural network potential and delta learning}

The neural network potential uses the graph network architecture available in TorchMD-Net~\cite{torchmdnet2}, which is an invariant message-passing neural network inspired by the SchNet and PhysNet architectures.
In this model, atomic environments are represented as nodes in a graph whose edges correspond to pairwise interactions within a cutoff radius.

The network uses inter-particle distances as geometric features, which are invariant to translations, rotations, and reflections.
These distances are used to generate continuous filter functions that modulate message-passing operations between neighboring nodes, similar to the interaction blocks used in SchNet.
Atomic embeddings are progressively updated through several graph convolution layers with residual connections, allowing the network to learn effective coarse-grained interaction potentials.

The network is the $\lambda$-repressor checkpoint from the dataset of Majewski \textit{et al.}~\cite{majewski2023}, with four interaction layers, an embedding dimension of 128, 18 radial basis functions, and lower and upper radial cutoffs of 3 and 12~\AA{} (at most 128 neighbors per bead), for a total of 370\,343 parameters.
The same checkpoint was used for all four proteins in every run reported here.
It serves as a fixed computational workload; we make no claim about its accuracy for any of the four proteins.

Delta learning was used to combine classical coarse-grained interactions with the neural network correction.
In this approach the total potential energy is expressed as
\[
E_{\mathrm{total}} = E_{\mathrm{classical}} + \Delta E_{\mathrm{NN}},
\] where $E_{\mathrm{classical}}$ corresponds to the bonded and excluded-volume interactions defined in the topology and $\Delta E_{\mathrm{NN}}$ is the energy correction predicted by the neural network.

\begin{figure*}
  \centering
  \includegraphics{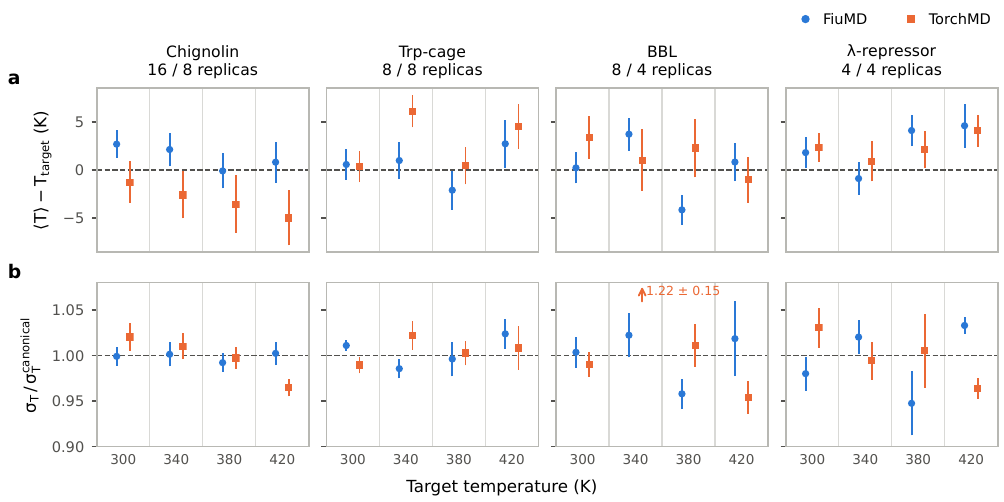}
  \caption{Kinetic temperature of the four proteins along a temperature ladder, FiuMD against TorchMD.(a)~Mean temperature minus the target, $\langle T \rangle - T_{\mathrm{target}}$;
the error is the standard error of block averages, taken as the largest value
over blocks of 10, 25 and 50~ps.
(b)~Standard deviation of $T$ over its canonical value,
$\sigma_T / \sigma_T^{\mathrm{canonical}}$;
the error comes from bootstrapping the replicas. Dashed lines mark the expected values. The first 50~ps were discarded, leaving 250 and 100~ps per replica,  details of the simulation in Tab.~\ref{tab:protocols}. The replicas of each engine are given above each column. The TorchMD point of BBL at 340~K ($1.22 \pm 0.15$) is off scale; it comes from a single temperature spike in one replica and drops to 1.04 without it.}
  \label{fig:temperature}
\end{figure*}

Coarse-grained bead identities were encoded through integer atom types that are passed to the neural network model as input identifiers.

\section{VALIDATION AND PERFORMANCE} \label{sec:tests}

We report four tests of the heterogeneous batch: forces and energies compared with isolated evaluations, deterministic trajectories compared with isolated runs, the kinetic-temperature distribution when the systems of one batch have different target temperatures, and the wall time of a mixed job.
The first two tests were run in double precision and the last two in single precision.
All of them combine the classical terms with the neural network potential as described in Table \ref{tab:protocols}.

\subsection{Forces and energies in a heterogeneous batch}

Figure~\ref{fig:forces} compares the total force on each system of the 60-system batch with five references.
A second evaluation of the same batch sets the noise floor of the comparison: over the three box sets, the relative difference has a median over systems of $2.1$--$2.2\times10^{-16}$ and a largest value of $5.5\times10^{-16}$.

Against the same system evaluated alone, the median relative difference is $2.1$--$2.3\times10^{-16}$ for isolated FiuMD and $2.9$--$3.1\times10^{-16}$ for isolated TorchMD.
The largest values over all systems and box sets are $6.6\times10^{-16}$ and $6.9\times10^{-16}$, respectively.
Permuting the order of the systems in the batch, or perturbing one other system, changes the force on a system by at most $5.6\times10^{-16}$ and $5.4\times10^{-16}$.
With the classical terms alone, the largest relative difference is $2.5\times10^{-16}$ against isolated FiuMD and $5.0\times10^{-16}$ against isolated TorchMD.

The potential energy of each system was compared in the same way, using the relative difference $|E-E_{\mathrm{ref}}|/\max(|E_{\mathrm{ref}}|, 1~\mathrm{kcal/mol})$.
It is at most $3.1\times10^{-15}$ in every comparison, with and without the neural network potential.

As a negative control, the four $\lambda$-repressor systems in the tight boxes were compared with isolated evaluations performed in the dataset box instead.
The relative force differences are then between 0.05 and 14, and the relative energy differences between 0.14 and 10.
For these systems the result depends on the box, and the comparison detects a wrong one.

In double precision, the forces and energies of a system inside a heterogeneous batch therefore agree with those of the isolated system at the level of a repeated evaluation, in all three box sets.

\subsection{Deterministic trajectories}

Figure~\ref{fig:nve} shows the divergence between batched and isolated NVE trajectories after 1000 steps. It was studied a batch of 60 systems described in Tab.~\ref{tab:systems} and Tab.~\ref{tab:protocols}.
The figure depicts the maximum atom displacement (\AA{}) for the four proteins studied. Each point is one system: the largest
displacement of any of its atoms between the batched run and
the same system run alone, with identical initial positions and
velocities, in FiuMD (circles) and in TorchMD (squares). 

All systems stay below $2\times10^{-11}$~\AA, almost three orders of magnitude under the $10^{-8}$~\AA{} tolerance of the test.

Similarly, Figure~\ref{fig:nve_time} compares trajectories started from identical positions and velocities and propagated for 1000 velocity Verlet steps, which corresponds to 1~ps for CLN and TRP and 0.5~ps for BBL and $\lambda$-repressor.
The separation between the batched and the isolated run is zero at the first step and grows gradually. At step 1000, the separation ranges from $3.4\times10^{-14}$ to $2.1\times10^{-11}$~\AA{} over all systems. The largest value at any step is $2.5\times10^{-11}$~\AA{}. All values are more than two orders of magnitude below the tolerance of $10^{-8}$~\AA{}.

Must-fail controls, a wrong time step and the same run without periodic images diverge by  $8 \cdot 10^{-3}$~\AA{} and  $10^{-5}$~\AA{} after a single step.

One system departs from the gradual growth. For one BBL system in the boxes with image contacts, the separation increases by more than an order of magnitude within about 30 steps near step 477, against both isolated references, and ends at $5.3\times10^{-12}$~\AA{} against FiuMD and $2.1\times10^{-11}$~\AA{} against TorchMD.
At that step one C$\alpha$--C$\alpha$--C$\alpha$ angle of this system reaches $179.9^\circ$.
The dihedral force is singular when three consecutive beads are collinear, and the force field used here has no angle term to prevent it, so round-off differences are amplified there in both engines.

This test covers 0.5 to 1~ps of deterministic dynamics.
It compares trajectories; it does not assess energy conservation or sampling.

\subsection{Kinetic temperature with per-system thermostat parameters}

Figure~\ref{fig:temperature} shows the kinetic temperature of the four proteins along a temperature ladder, FiuMD (circles symbols) against TorchMD (squares symbols). Fig.\ref{fig:temperature}a shows the mean temperature minus the target $\langle T \rangle - T_{\mathrm{target}}$, the error bars are the standard error of block averages, calculated as the largest value over blocks of 10, 25 and 50 ps. 
And, Fig.~\ref{fig:temperature}b depicts the standard deviation of $T$ over its canonical value $\sigma_T / \sigma_T^{\mathrm{canonical}}$, with $\sigma_T^{\mathrm{canonical}} = T_{\mathrm{target}}\sqrt{2/3N}$ for $N$ beads, and the error comes from bootstrapping the replicas. In both cases, the dashed lines represent the expected values. The first 50~ps were discarded, leaving 250 and 100~ps per replica, more  details of the simulation are in Tab.~\ref{tab:protocols}.

Each FiuMD batch held the four proteins at four different target temperatures (300, 340, 380 and 420 K). Across four runs the temperatures were rotated. TorchMD has a single thermostat temperature per run, each protein–temperature pair required a separate simulation.

The mean kinetic temperature matched the target within $\approx 4.5$ K in FiuMD and $\approx 6$ K in TorchMD. Sharing a batch with systems at other temperatures does not bias the temperature of a system: each one relaxes to its own target 

The width of the temperature distribution $\sigma_T$, agrees with the canonical value within 0.95-1.03 for FiuMD. In TorchMD we observed a single off-scale point for BBL at 340 K which comes from a single transient spike. 

FiuMD and TorchMD give statistically indistinguishable temperatures: in every one of the 16 protein–temperature pairs the two mean temperatures differ by at most 2.0 combined standard errors. Thermostatting heterogeneous systems at different targets in one batch therefore reproduces the canonical temperature distribution obtained by running each system alone. 

FiuMD correctly samples the canonical ensemble for each system of a heterogeneous batch. 

\subsection{Wall time of a mixed job}

\begin{figure}[!htbp]
    \centering
    \includegraphics[width=\columnwidth]{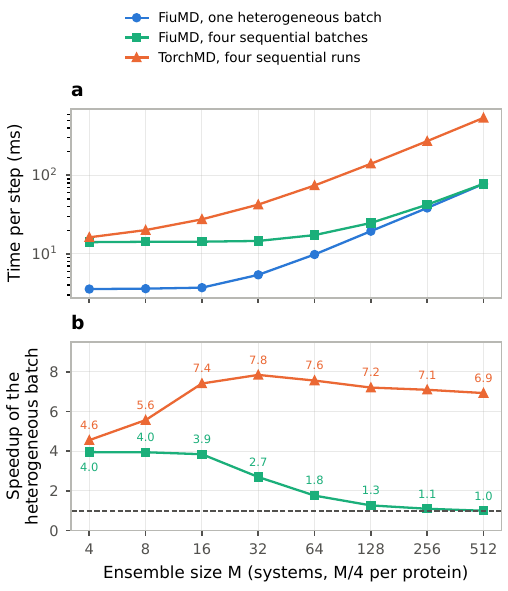}
    \caption{Wall time of a mixed job. (a)~Wall time per integration step: median of 5 repeats of 100 steps each, after 20 warm-up steps. (b)~Speedup of the heterogeneous batch, $t_{\mathrm{sequential}} /
t_{\mathrm{heterogeneous}}$, over each baseline; the dashed line marks equal time.    
    An ensemble of $M$ systems, $M/4$ of each
of the four proteins, three cases are compared: one heterogeneous FiuMD batch (blue circle) and four sequential batchs (one per protein) in FiuMD (green squared) and 
TorchMD (orange triangle); the time of the sequential baselines is the sum of the four jobs. 
Every step evaluates the priors and the neural network potential in single precision and applies a Langevin thermostat. All timings on one NVIDIA GeForce RTX 5060 Ti. Values and peak GPU memory are listed in Table~\ref{tab:timing}}
    \label{fig:timing}
\end{figure}

Figure~\ref{fig:timing} shows the wall time per step of a job containing $M/4$ replicas of each of the four proteins, run in three ways.
One heterogeneous FiuMD batch takes 3.6~ms per step at $M=4$ and 77.8~ms at $M=512$; four homogeneous FiuMD batches run one after another take 14.1~ms and 77.9~ms.
The five repeated measurements of each total differ from their median by at most 2.4\%.

Four TorchMD runs, one per protein and executed one after another, take 16.2~ms per step at $M=4$ and 539~ms at $M=512$.
The heterogeneous batch is therefore 4.6 times faster at $M=4$, 7.8 times faster at $M=32$, where the ratio is largest, and 6.9 times faster at $M=512$.
At $M=512$ and a timestep of 1~fs, this corresponds to an aggregate throughput, summed over the 512 systems, of 569~ns/day (about 1.1~ns/day per system) against 82~ns/day.

The comparison with four homogeneous FiuMD batches, also executed one after another, separates the effect of mixing proteins from the effect of batching replicas.
The heterogeneous batch is 4.0 times faster at $M=4$ and $M=8$, 2.7 times faster at $M=32$, 1.3 times faster at $M=128$, and equally fast (ratio 1.00) at $M=512$.
Up to $M=16$ the time per step of the heterogeneous batch is nearly independent of $M$ (3.6 to 3.7~ms), and each of the four homogeneous batches takes about the same time as the whole heterogeneous batch.
Merging the four batches into one then removes about three quarters of the cost.
For larger $M$ the time per step grows with $M$ and this advantage disappears.

At $M=512$ most of the gain over TorchMD therefore comes from batching the replicas of each protein, not from mixing proteins.
The four homogeneous FiuMD batches of 128 replicas are 31.5, 17.3, 6.0, and 3.6 times faster than the corresponding TorchMD runs for CLN, TRP, BBL, and $\lambda$-repressor, respectively.

Table~\ref{tab:timing} lists the median times, ratios, and peak-memory ratios for every batch size.
\begin{table}[!h]
\caption{\label{tab:timing}Median wall time per step (ms) for $M/4$ replicas of each of the four proteins, in single precision on one RTX~5060~Ti. Sequential times are sums of the four species-specific medians. $S_{\rm F}$ and $S_{\rm T}$ divide the sequential FiuMD and TorchMD times by the mixed FiuMD time. $R_{\rm mem}$ is the peak GPU memory of the mixed batch divided by the largest peak among the four homogeneous FiuMD batches.}
\begin{ruledtabular}
\begin{tabular}{rrrrrrr}
$M$ & Mixed & 4$\times$FiuMD & 4$\times$TorchMD & $S_{\rm F}$ & $S_{\rm T}$ & $R_{\rm mem}$ \\
\hline
4 & 3.56 & 14.08 & 16.24 & 3.96 & 4.6 & 1.15 \\
8 & 3.60 & 14.24 & 20.07 & 3.95 & 5.6 & 1.28 \\
16 & 3.71 & 14.28 & 27.50 & 3.85 & 7.4 & 1.37 \\
32 & 5.40 & 14.59 & 42.38 & 2.70 & 7.8 & 1.60 \\
64 & 9.81 & 17.32 & 74.22 & 1.77 & 7.6 & 1.60 \\
128 & 19.45 & 24.72 & 140.26 & 1.27 & 7.2 & 1.64 \\
256 & 38.40 & 42.31 & 272.63 & 1.10 & 7.1 & 1.69 \\
512 & 77.77 & 77.86 & 539.26 & 1.00 & 6.9 & 1.71 \\
\end{tabular}
\end{ruledtabular}
\end{table}

The heterogeneous batch holds all systems in memory at once.
At $M=512$ its peak memory allocated by PyTorch is 4.1~GiB, 1.7 times that of the largest of the four homogeneous batches.

These timings were measured in single precision on one GPU and for one neural network model.
The equivalence tests above were run in double precision; we report no equivalence test in single precision.
We did not profile the GPU, so we do not attribute the measured times to a specific mechanism.

\section{Conclusions and Outlook}
FiuMD enables multiple independent coarse-grained protein simulations with different system sizes, topologies, force-field parameters, periodic boxes, timesteps, and thermostat parameters to be advanced within a single GPU batch. Numerical validation demonstrates that this heterogeneous batching preserves the behavior of isolated simulations: forces and energies agree at the level of repeated evaluations, while deterministic trajectories remain within $2.5\times10^{-11}$~\AA{} of their isolated counterparts over 1000 steps. Additional tests with different target temperatures confirm that each system maintains its own thermostat and correctly reproduces the canonical kinetic temperature distribution. Negative controls further demonstrate that the validation framework is sensitive to errors arising from incorrect box dimensions, timesteps, or periodic image handling.

Performance benchmarks show that, for a mixed job of four protein systems on a single GPU, heterogeneous batching achieves a 4.6–7.8-fold speedup per step compared with sequential TorchMD simulations. It is also up to 4.0 times faster than running four separate homogeneous FiuMD batches for smaller batch sizes, while achieving comparable performance for larger batches. These results highlight the potential of FiuMD to improve GPU utilization and computational throughput when simulating collections of heterogeneous molecular systems.

The current implementation has been validated using C$\alpha$ protein models with bond, repulsive $r^{-6}$, and dihedral interactions in orthorhombic periodic boxes, together with a single TorchMD-Net model on one GPU. 
While these results establish the feasibility and numerical reliability of the approach, they also provide a starting point for further development. 
Future work will focus on expanding the range of supported interaction potentials and simulation features, improving computational and memory efficiency, and extending validation to additional molecular systems and simulation conditions. 
Further studies will also assess the impact of heterogeneous batching on conformational sampling and simulation performance across larger and more diverse workloads.

Overall, FiuMD provides a flexible foundation for combining heterogeneous molecular simulations on GPUs, and its modular approach offers opportunities for continued development and integration of new capabilities.

\section*{Acknowledgments}

We thank the developers of TorchMD and TorchMD-Net for making their code openly available, and Majewski \textit{et al.} for the coarse-grained protein data and models used in this work.
We thank the Grupo Vinculado Bionanotecnolog\'ia y Sistemas Complejos (INFAP, CONICET--UNSL--UTN) and the Instituto de Fisica del Sur (IFISUR, CONICET--UNS) for providing the computational resources used in this work, and the Facultad Regional San Rafael of the Universidad Tecnol\'ogica Nacional (UTN) for the space and facilities that host them.
P.B.T. acknowledges a postdoctoral fellowship from CONICET, and H.D.H.R. and L.N. acknowledge doctoral fellowships from CONICET.

\section*{Author contributions}

\textbf{Hugo D. Hern\'andez R.}: Conceptualization, Methodology, Software, Visualization, Writing -- original draft, review \& editing. \\
\textbf{Paola B. Torres}: Methodology, Software, Validation, Formal analysis, Visualization, Writing -- original draft, review \& editing.\\
\textbf{Lara Negri}: Formal analysis, Writing -- original draft, review \& editing.\\
\textbf{Roc\'io L. Cecchini}: Supervision, Writing -- review \& editing.\\
\textbf{Nicol\'as A. Garc\'ia}: Supervision, Funding acquisition, Writing -- review \& editing.\\
All authors reviewed and approved the final manuscript.

\section*{Code availability}

FiuMD is an MIT-licensed fork of TorchMD~\cite{torchmd}, available at \url{https://github.com/hernandezhdd/fiumd} and archived on Zenodo as release v0.1.0 (\href{https://doi.org/10.5281/zenodo.24233116}{doi:10.5281/zenodo.24233116}).
The library code of that release is identical to revision \texttt{433cedf}, which was used for all simulations in this work.
It was tested with Python~3.10.18, PyTorch~2.13.0+cu130, CUDA~13.0, TorchMD~1.1.2 and TorchMD-Net~3.0.3; after installing these prerequisites and MoleculeKit, run \texttt{pip install -e .} from the repository root.
The folder \texttt{paper/} of the release contains the simulation drivers, the figure scripts, a script that downloads the inputs and the exact command behind each figure.
The topologies, prior force field, starting structures and network checkpoint come from the dataset of Majewski \textit{et al.}~\cite{majewski2023}.
\texttt{examples/two\_proteins.py} advances chignolin and Trp-cage in one batch at two temperatures.

\section*{Data availability}

The coarse-grained topologies, prior force-field parameters, starting structures and network checkpoint were taken from the dataset of Majewski \textit{et al.}~\cite{majewski2023}; 
the script \texttt{paper/datos/download\_data.py} of the FiuMD v0.1.0 release retrieves them and writes the exact input files used here. 
The raw outputs behind Figs.~\ref{fig:forces}--\ref{fig:timing} are archived on Zenodo at
\href{https://doi.org/10.5281/zenodo.24236761}{doi:10.5281/zenodo.24236761}, 
together with the commands that rebuild each figure.
Full trajectories are available from the corresponding author upon reasonable request.

\bibliography{fiumd}

\end{document}